---------------------------------------------------------------

\documentstyle[aps,prl]{revtex}
\input{psfig}
\def\salmon{{\it Salmonella typhimurium }}
\def\bs{{\it Bacillus subtilis }}
\twocolumn
\begin{document}
\input{epsf}
\title {AGGREGATION PATTERNS IN STRESSED BACTERIA }
\author{Lev Tsimring and Herbert Levine}
\address{ Institute for Nonlinear Science
University of California, San Diego La Jolla, CA  92093-0402}
\author{Igor Aranson}
\address{Dept. of Physics, Bar-Ilan University, Ramat Gan, ISRAEL}
\author {Eshel Ben-Jacob, Inon Cohen and Ofer Shochet}
\address{School of Physics and Astronomy,
Raymond \& Beverly Sackler Faculty of Exact Sciences,
Tel-Aviv University, Tel-aviv 69978, ISRAEL}
\author{William N. Reynolds}
\address {Complex Systems Div., Los Alamos Natl. Lab.  Los Alamos, N.M.}

\maketitle
\begin{abstract}
We study the
formation of spot patterns seen in a variety of bacterial species
when the bacteria are subjected to oxidative stress
due to hazardous byproducts of respiration. Our approach consists of
coupling the cell density field to a chemoattractant concentration as
well as to nutrient and waste fields. The latter serves as a triggering
field for emission of chemoattractant.  Important elements in the proposed
model
include the
propagation of a front of motile bacteria radially outward form an
initial site, a Turing instability of the uniformly dense state and a
reduction of motility for cells sufficiently far behind the front.
The wide variety of patterns seen in the experiments is explained as
being due the variation of the details of the initiation of the
chemoattractant emission as well as the transition to a
non-motile phase.
\end{abstract}

Over the past few years, there has been a significant increase in our
understanding of how spatial patterns emerge via the
propogating interfacial dynamics of
non-equilibrium systems\cite{kessler,benjacob}. We have discovered
how seemingly different systems can nonetheless exhibit strikingly
similar behavior, due to the existence of nonequilibrium pattern selection
principles. These principles rely on the idea that the final
structure which emerges from an initially (linearly) unstable state is
affected mostly by the nature of the instability, the possible existence
of stable, highly nonequilibrium isolated steady-state structures (such as the
single dendrite in nonequilibrium solidification), and the competition
between globally ordererd arrangements of these structures as compared to
more disordered morphologies.
This framework has been applied
to a variety of systems of physical, chemical and most
recently\cite{biology} biological interest.

In this work\cite{Nature}, we analyze spot and stripe patterns seen
in bacterial growth experiments; the first such results were due to
Berg and Budrene\cite{Budrene} in {\em e. coli}.
They found that cells could aggregate chemotactically,
resulting in a wide variety of different colony structures ranging from arrays
of spots to
radially oriented stripes to arrangements of more complex elongated
spots. In
their study, {\em e. coli} were grown on single carbon source
media and the appearance of patterns only in the case of highly
oxidized nutrient suggested that respiratory by-products (leading
to oxidative stress\cite{stress})
trigger the observed chemotactic behavior.
Two of their figures are reproduced in Fig.1. It is worth noting that this
class of patterns is not limited to {\em e. coli}; similar structures have
been seen in \salmon \cite{Blat} and in \bs \cite{ofer}. We suspect
that this class of patterns is a universal ``possibility" for microbial
systems aggregating in the face of adversity.

Normally, bacteria divide and spread out into regions of
initially low density.  This can take the form of an expanding
circle\cite{shapiro}, or if some metabolic factor is in short supply, a
(set of) expanding ring(s)\cite{rings}. In the presence of the
aforementioned oxidative stress, the bacteria begin to emit a chemoattractant
(believed to be aspartate) which causes them to aggregate via biasing
their motion. Eventually, the bacteria turn non-motile,
freezing the pattern into place.  From the modeling point of view, the coupling
of the
chemoattractant diffusion equation to the bacterial density evolution
leads to a Turing-like instability of the uniform density state. If we
were to start the entire system at uniform bacterial density,
patterns would develop as soon as the concentration of
waste reaches a threshold value. These patterns
would be defect-ridden, governed by the precise details of the
initial conditions. This is indeed what was observed in the
case of purposeful addition of hydrogen peroxide.  The
symmetric structures seen during growth from a
single inoculated site are thus due to the interplay of the expansion
with the Turing instability as triggered by the waste field, and possibly
autocatalytically by the attractant field itself.  Below, we will
see how this occurs within our model.

Based on the above, we propose the following set of continuum equations
for this system\cite{bruno}:
\begin{eqnarray}
\dot{\rho} & = D_{\rho} &\vec{\nabla}^2 \rho +
G(\rho ,n) - v_c \vec{\nabla} \cdot (\rho \vec{\nabla} c )
- I[w]\rho\nonumber \\
\dot{\rho_n} & = & I[w]\rho\nonumber\\
\dot{n} & = & \ D_n \vec{\nabla} ^2 n - \alpha \rho n \nonumber \\
\dot{w} & = & \ D_w \vec{\nabla} ^2 w + \alpha_w \rho n -\beta_w w-\gamma \rho
w
\nonumber \\
\dot{c} & = & \ D_c \vec{\nabla} ^2 c + T[w,c]\rho - \beta_c c \label{1} \\
\end{eqnarray}
Here $\rho$ and $\rho_n$ are densities of motile and non-motile bacteria,
$w$ is a concentration of respiratory waste products,
$c$ is the chemoattractant concentration which emitted by the bacterium,
and $n$ is nutrient which is eaten by bacteria. The functionals
$T$ and $I$ are thresholds to be discussed below.
We have used the freedom to rescale to set some coefficients to
unity.

This equation
contains a  standard growth term which we typically take to be of the form $G \
=\
r\rho^{p} n/(n+n_0)-\rho^3$, with $p \simeq 1$ \cite{numerics};
this form reflects the nutrient-inhibited
growth of cells
at low food concentration and finite reproduction rate which is achieved in
a nutrient-``rich" limit.  At high density of bacteria the growth is limited
by the nonlinear term $-\rho^3$. Motion is governed by diffusion and by a
chemotactic term representing the response of the bacteria to a
gradient in the attractant. As already mentioned, it will be important to
incorporate the fact that sufficiently far behind the advancing front,
the bacteria differentiate into a non-motile form. We assume that the
transition occurs due to the accumulation of effects due to
starvation; specifically, the transition
to the non-motile phase occurs when $n_c\equiv\int_0^t\, (n_0-n)\theta(n_0-n)\,
dt$ exceeds some threshold value $n_{tr}$, i.e.  by setting
$I[w]=\delta \Theta(n_c-n_{tr})$\cite{lagrangian}
($\Theta(x)$ is the  Heaviside function).
The rate of waste accumulation is proportional to the nutrient consumption
term, and we assume that the
waste decomposes with a rate dependent on the bacterial
density, as this is in fact the underlying reason for the bacteria to
aggregate. Finally, the
emission of chemoattractant $c$ is proportional to the local
density of bacteria, and it is triggered by
local waste field;  specifically, we assume that the chemoattractant is
emitted if either $w>w_0 \ \mbox{and} \  c>c_0$ or $w>w_1$, where $w_1>w_0$
(formally, $T[w,c]=\mbox{sign}(\Theta(w-w_0)\Theta(c-c_0)+\Theta(w-w_1))$).
The difference between $w_1$ and $w_0$ represents
possible autocatalytic behavior of the attractant.

To understand the structure of our model, it is convenient to first
consider the case of constant uniform nutrient ($n \gg n_0,\ \alpha=0$)
and in the absence of any waste effects (i.e. $T \ = \ 1$). The system
now has an unstable
steady-state $\rho = c = 0$. If we start with an initial condition of
localized bacteria density amidst a sea of $\rho = 0$, the density will
spread. In the absence of coupling to $c$, the front would move at a
speed at $2 \sqrt{D_\rho r}$ via the usual marginal stability
criterion\cite{vansaarlos}. By continuity, the front will continue to
expand as long as $v_c$ is not too large.
On the other hand, the non-trivial uniform state is given by $\rho = \sqrt{r},
c=1/\beta$.
The stability of this state to perturbations with wave
vector $q$ is given by the roots of
\begin{equation} (\omega + D_{\rho}
q^2 +2\ r)(\omega +D_c q^2 +\beta_c) = v_c r^{1/2}q^2
\end{equation}
It is easy to see that for large enough chemotactic response $v_c$
this system has a band of wavevectors with purely real
and positive growth rates $\omega$. This is just the Turing instability
described above. It does not require different diagonal diffusivities
$D_{\rho}$ vs. $D_c$ but instead relies on there being a large cross
diffusivity. Combining these arguments
leads one to expect that generically there exists an intermediate range
of $v_c$ for which the bacteria density will propagate outward and
create an expanding Turing-unstable region.
Adding the nutrient field back in does not alter the above conclusions
in any qualitatively important manner.  The importance of waste dynamics
will be discussed below.

To simulate the above equations, we used a split-step
spectral code on a 128x128 lattice; specifically, the linear parts of
the evolution were treated by FFT and the nonlinear pieces by explicit
finite differences. Lattice anisotropy which
comes primarily from the discretizing the nonlinear diffusion terms in
eq. (\ref{1})  was minimized by computing them simultaneously on two
lattices which differ by 45$^{o}$ rotation\cite{lattice}.
Also, lattice discretization effects can lead to spurious (slightly)
negative values of the density $\rho$; this is presumably due to the
lattice trying to mimic the possibility in the continuum of having
precisely localized structures due to the nonlinear
diffusivity\cite{nonlinear}. To deal with this
we used $p>1$ (typically $\approx1.5$) and indeed it greatly reduced
this spurious effect without  any qualitative change in
the dynamics. We can immediately verify the above simple
features and proceed to discuss the formation of complex
ordered structures.

Let us first focus on colony dynamics in the absence of a
threshold for chemoattractant emission($T(w,c)\equiv 1$).
In Figure 2, we show a snapshot of a typical simulation of the
model equations.  As the interface propagates outwards, it creates a
set of concentric rings.  The breakup of the rings into spots
occurs somewhere behind the front, dependent on the
specific value of $v_c$;  for small enough $v_c$, the rings
do not break up. Ordered arrangements of the spots are
hard to achieve, since the interaction
between different rings is extremely small. Moreover, the appearance of
a number of rings behind the outer rings is not in accord
with the Budrene and Berg experiment. In most cases the spots appeared
right behind the outer ring.

To account for this effect, we impose the threshold for
chemoattractant emission. The existence of this
strong nonlinearity makes the rear
part of the outer ring extremely sensitive to the structure
of the waste and chemical fields. In particular, the
emission is triggered preferentially at certain angular
positions as opposed to along an entire ring. This then
can cure the two previous problems -  proper thresholding
can lead directly to spots and previous spots strongly
bias the formation of new spots. In fig. 3, we show the
results of full model simulations, showing significantly
improved agreement with experiment. We now discuss in
turn the threshold choices that lead to the differing
results in figs 3.

In  figure 3a, we see that spots form at positions
such that they will eventually form radial rows; these rows are
quite similar to those in the
experimental pattern in Fig. 1a. The reason for this is that the
concentration of $c$ in the ring of currently aggregating
spots is nonuniform
and due to diffusion this inhomogeneity spreads into the rear part of the
outer ring. This biases the location for the new spots so they are
created right in front of the previous spots.
Once the spots form, they are not allowed to move very far
since as nutrient depletes, bacteria transform into a
non-motile phase. If the chemotactic response ($v_c$) is reduced,
instead of radially aligned spots radial stripes are formed;
this too has been seen experimentally (see Fig.1b and 3b).
We note
in passing that the brightness of spots in experimental images is
greater for motile bacteria so in all numerical figures we show the gray scale
plots of $(\rho+0.5\rho_n)$.

In the other main experimental spot pattern, the spots arrange themselves in
a manner such that the new spot appears roughly between two
pre-existing ones at the previous ring.
By adjusting the values of $c_0$ and $w_1$
we can produce a similar pattern as well (see fig. 3c).
By significant increase of the threshold $c_0$ we
effectively made the residual level of chemoattractant
irrelevant, and instead the waste
threshold $w_1$ plays the major role.  We should note, however,
that this structure
seems to be missing the visually striking spirals which
create the sunflower impression in the original BB figure
(fig 1a. of their paper). Whether this
difference represents a shortcoming of our model which tends
to keep spots along circular rings, or, as we feel more
likely, that a larger system with less anisotropy and perhaps a
different set of parameters could reproduce this visual impression is a
question for future work.

To summarize, we have shown how the interplay of
front propagation and a Turing-type instability can lead to spot patterns
similar to those observed in bacterial aggregation. We have
concentrated on generic mechanisms which should remain important
regardless of the details of the explicit biological interactions as
these details become clearer. One important conclusion is that
the biological mechanism for the ``turning on" of chemotattractant emission
is of critical importance in giving rise to the observed colonies.
Of course, attempts at more quantitative
comparison would require as input the functional dependencies of all
the various pieces entering into the dynamics. We feel that this level
of detailed modeling is a reasonable undertaking only after a
demonstration that physics does indeed have a hope of describing what
is going on without the need to invoke an incredibly complex hierarchy
of biological mechanisms. This paper has been attempt to demonstrate
exactly this point and to thereby provide motivation for a future
quantitative study.

\begin{figure}
\caption{Two examples of patterns of bacteria \protect{\em e. coli}
in experiments by
Budrene and Berg [1]: radial alignment of spots (a), radially oriented
stripes (b)}
\label{fig1}
\end{figure}
\begin{figure}
\caption{
A snapshot of bacteria density ($\rho+0.5\rho_n$) at $t=18$
within the model  with $D_{n}=0.4,\ D_c=0.2,\
D_w=0.4,$ $\alpha=0.12,$ $r=1.8,$ $\beta_c=1$, $\alpha_w=0.3$, $\beta_w=0.1$,
$\gamma=0.2$, $n_0=0.5,\ n_{tr}=0.03,\, \delta=0.5, \ v_c=6.0,$,
thresholds for chemoattractant disabled, system size is 40,
time step 0.02. Concentric rings are formed in the wake of the outer ring}
\end{figure}
\begin{figure}
\caption{
{\em (a)}
Same as in Figure 2
but with enabled thresholds ($w_0=0.3, w_1=0.55, c_0=0.04$) and $v_c=7$; radial
rows of spots
are clearly seen (cf. Figure 1a);
{\em (b)}
Same as in {\em a} but with weaker chemotaxis ($v_c=6$), radial stripes
similar to Figure 1b are seen;
{\em (c)} Pattern of $\rho+0.5\rho_n$ within the model (\protect\ref{1})
with different thresholds ($w_0=0.3, w_1=0.5,c_0=0.5$) and higher
chemotaxis ($v_c=9$), other parameters are the same as in {\em b}.
A staggered alignment is seen far from the center.}
\label{fig3}
\end{figure}


\begin{thebibliography}{99}


\bibitem{kessler}  D. Kessler, J. Koplik and H. Levine,
{\em Adv. in Phys.} {\bf 37}, 255 (1988).

\bibitem{benjacob}  E. Ben-Jacob
and P. Garik, {\em Nature} {\bf 343}, 523 (1990).


\bibitem{biology} E. Ben-Jacob, et. al, Nature {\bf
368}, 46 (1994); D. Kessler and H. Levine, Phys. Rev. {\bf E48}, 4801
(1993).

\bibitem{Nature} Some of our ideas have appeared in
in Nature, scientific correspondence. (in press).

\bibitem{Budrene} E.Budrene, H.Berg, Nature,{\bf 349}, 630 (1991).

\bibitem{stress} G. Storz, L. A. Tartaglia, S. B. Farr, and B. N. Ames,
Trends Genet. {\bf 6}, 363 (1990).

\bibitem{Blat} Y. Blat and M. Eisenbach. {\em J. Bacteriology}, in press.

\bibitem{ofer}  E. Ben-Jacob and O. Shochet, unpublished.

\bibitem{shapiro} J. Shapiro, Physica {\bf D49}, 214 (1991).

\bibitem{rings} J. Adler, Science {\bf 153}, 708 (1966); R. Nossal,
Exp. Cell Res. {\bf 75}, 138 (1972); A. Wolfe and H. Berg, Proc. Nat.
Acad. USA, {\bf 86}, 6973 (1989).

\bibitem{bruno} The only previous work to date on this system used
a simple model without thresholds; W. J. Bruno, Los Alamos
CNLS newsletter, {\bf 82}, 1 (1992) and private communication.

\bibitem{numerics} The idea of using $p>1$ to account for the discrete
character of an aggregating particle (here a bacterium) arose in the context
of diffusion-limited aggregation; see E. Brener, H. Levine and Y. Tu,
Phys. Rev. Lett. {\bf 66}, 1978 (1991).

\bibitem{lagrangian} Strictly speaking, the integrated nutrient deficit
should be done in "Lagrangian" variables, i.e. finding the nutrient deficit for
individual bacteria as they move through space. This is hard to implement in
our "Eulerian" description and in any case should be a small effect since
in our model bacteria do not move over distances over which there are
significant
changes in nutrient concentration.

\bibitem{vansaarlos} W. Van Saarloos, Phys. Rev. {\bf A39}, 6327 (1989)
and references therein. This actually applies only to the case $p=1$, but
as long as $p$ is not too large, the selected velocity changes only slightly
from the marginal stability prediction.

\bibitem{lattice} The effect of residual anisotropy was determined by letting
the
system evolve for some time, rotating the fields by an irrational angle, and
then continuing the time evolution. This procedure did not interfere with
the radial row structure discussed later in the text.

\bibitem{nonlinear} For some results regarding nonlinear diffusion
equations, see G. I. Barenblatt, {\em Similarity, self-similarity, and
intermediate asymptotics } Consultants Bureau, New York (1979).

\end{thebibliography}
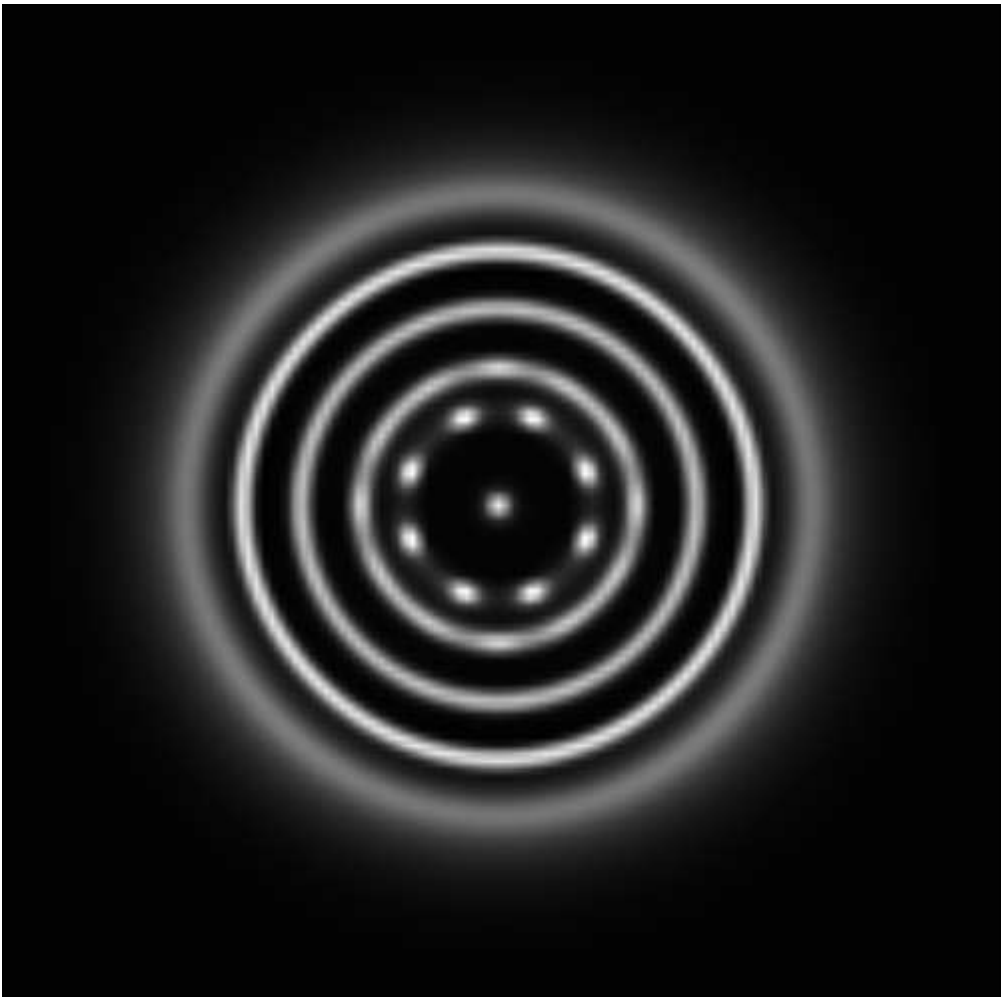
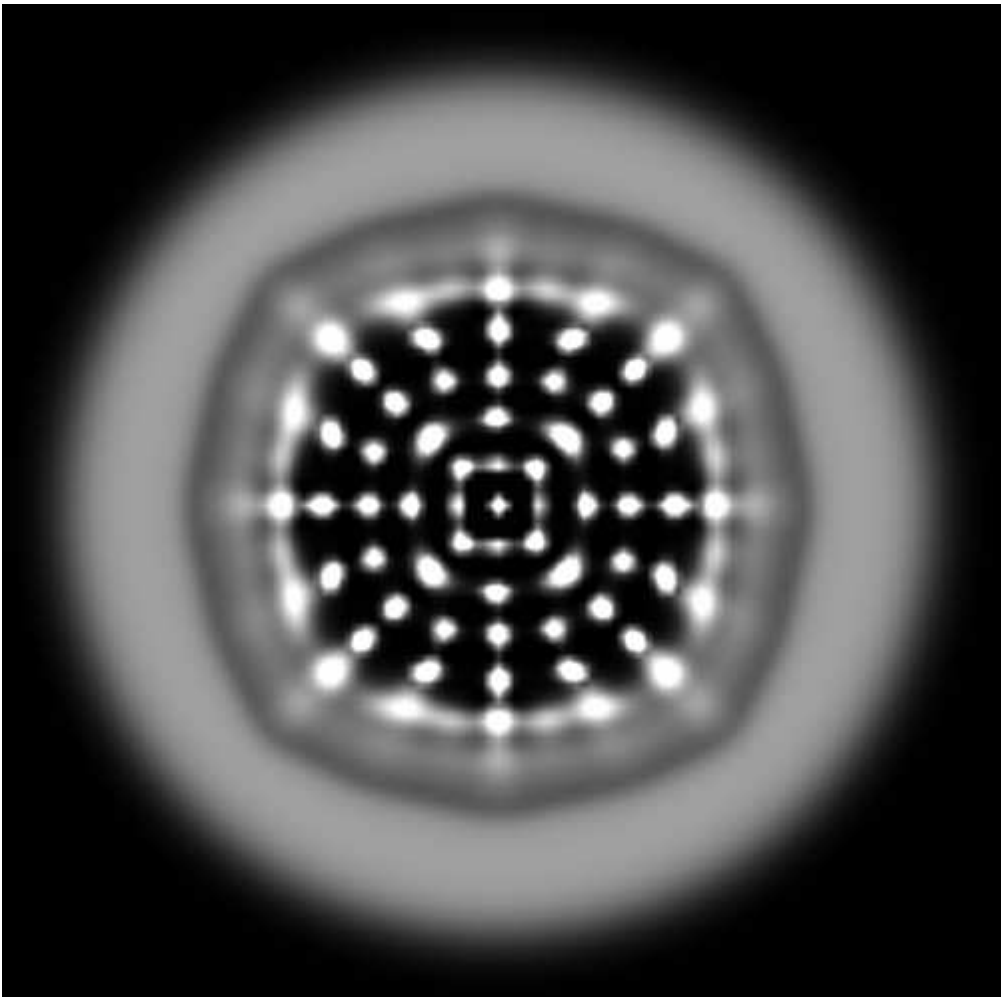
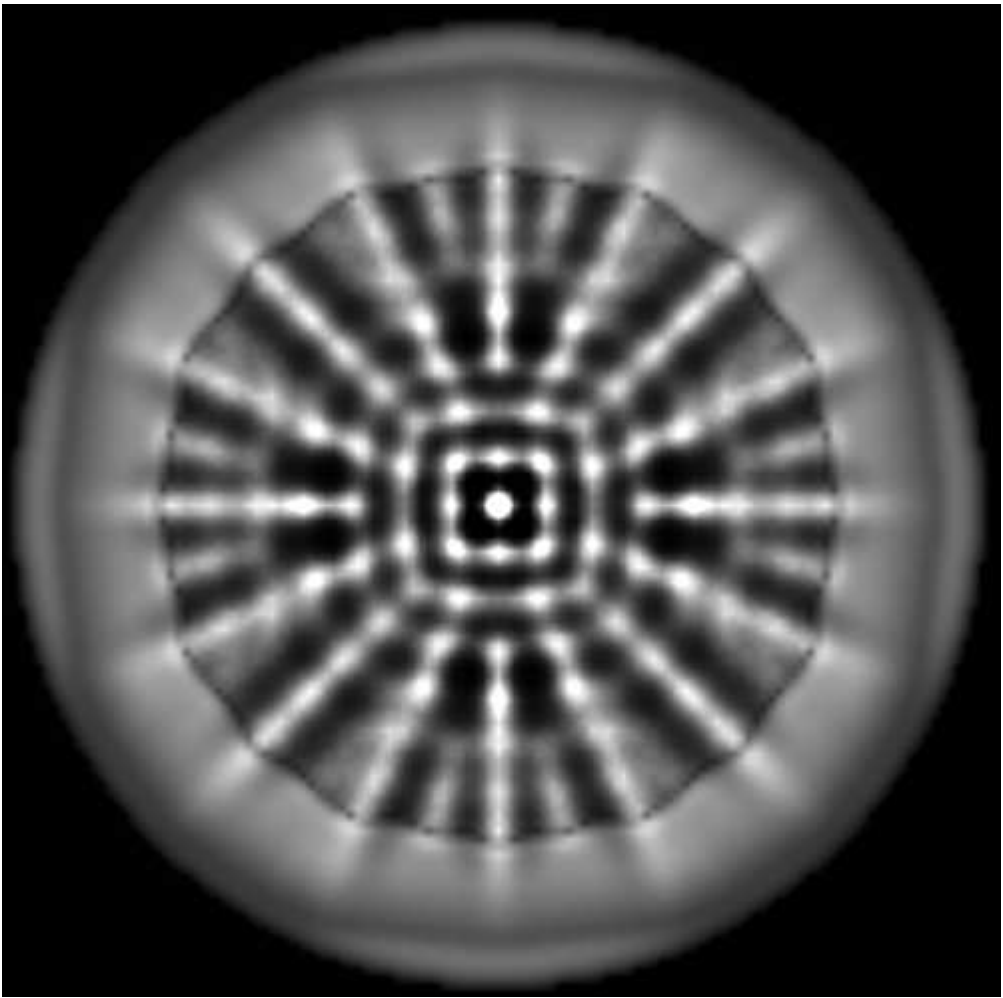
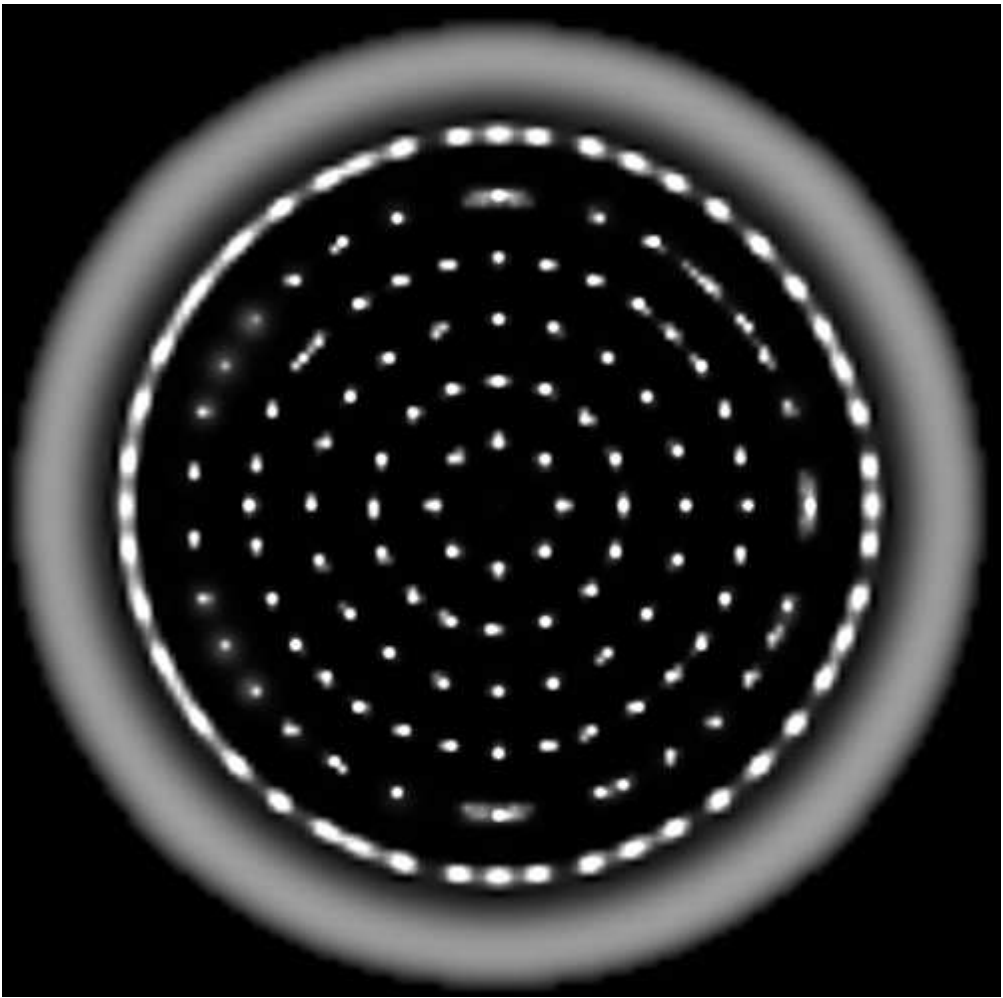
\end{document}